\documentclass[cits]{PoS}
\usepackage{xspace}

\newcommand{\Al}{$^{26}$Al\xspace}

\newcommand{\Co}{$^{56}$Co\xspace} 

\newcommand{\Ni}{$^{56}$Ni\xspace}
\newcommand{\Ti}{$^{44}$Ti\xspace}
\newcommand{\Sc}{$^{44}$Sc\xspace}

\newcommand{\gam}{\ensuremath{\gamma}}

\newcommand{\cms}{cm\ensuremath{^{-2}} s\ensuremath{^{-1}}\xspace}

\let\jnl=\rmfamily
\def\refe@jnl#1{{\jnl#1}}%

\newcommand\aj{\refe@jnl{AJ}}%
\newcommand\actaa{\refe@jnl{Acta Astron.}}%
\newcommand\araa{\refe@jnl{ARA\&A}}%
\newcommand\apj{\refe@jnl{ApJ}}%
\newcommand\apjl{\refe@jnl{ApJ}}%
\newcommand\apjs{\refe@jnl{ApJS}}%
\newcommand\ao{\refe@jnl{Appl.~Opt.}}%
\newcommand\apss{\refe@jnl{Ap\&SS}}%
\newcommand\aap{\refe@jnl{A\&A}}%
\newcommand\aapr{\refe@jnl{A\&A~Rev.}}%
\newcommand\aaps{\refe@jnl{A\&AS}}%
\newcommand\azh{\refe@jnl{AZh}}%
\newcommand\memras{\refe@jnl{MmRAS}}%
\newcommand\mnras{\refe@jnl{MNRAS}}%
\newcommand\na{\refe@jnl{New A}}%
\newcommand\nar{\refe@jnl{New A Rev.}}%
\newcommand\pra{\refe@jnl{Phys.~Rev.~A}}%
\newcommand\prb{\refe@jnl{Phys.~Rev.~B}}%
\newcommand\prc{\refe@jnl{Phys.~Rev.~C}}%
\newcommand\prd{\refe@jnl{Phys.~Rev.~D}}%
\newcommand\pre{\refe@jnl{Phys.~Rev.~E}}%
\newcommand\prl{\refe@jnl{Phys.~Rev.~Lett.}}%
\newcommand\pasa{\refe@jnl{PASA}}%
\newcommand\pasp{\refe@jnl{PASP}}%
\newcommand\pasj{\refe@jnl{PASJ}}%
\newcommand\skytel{\refe@jnl{S\&T}}%
\newcommand\solphys{\refe@jnl{Sol.~Phys.}}%
\newcommand\sovast{\refe@jnl{Soviet~Ast.}}%
\newcommand\ssr{\refe@jnl{Space~Sci.~Rev.}}%
\newcommand\nat{\refe@jnl{Nature}}%
\newcommand\iaucirc{\refe@jnl{IAU~Circ.}}%
\newcommand\aplett{\refe@jnl{Astrophys.~Lett.}}%
\newcommand\apspr{\refe@jnl{Astrophys.~Space~Phys.~Res.}}%
\newcommand\nphysa{\refe@jnl{Nucl.~Phys.~A}}%
\newcommand\physrep{\refe@jnl{Phys.~Rep.}}%
\newcommand\procspie{\refe@jnl{Proc.~SPIE}}%

\title{Gamma-Rays from Positron Annihilation}

\ShortTitle{Positron Annihilation}

\author{\speaker{Roland Diehl}\\
        Max Planck Institut f\"ur extraterrestrische Physik, D-85748 Garching, Germany\\
        E-mail: \email{rod@mpe.mpg.de}}
\author{Mark Leising\\
        Clemson University, Clemson SC 29631 , USA\\
        E-mail: \email{lmark@clemson.edu}}

\abstract{SPI on INTEGRAL has provided spectra and a map of the sky in the emission from annihilations of positrons in the interstellar medium of our Galaxy. From high-resolution spectra we learned that a warm, partially-ionized medium is the site where the observed gamma-rays originate. The gamma-ray emission map shows a major puzzle for broader astrophysics topics, as it is dominated by a bright and extended apparently spherical emission region centered in the Galaxy's center. Only recently has the disk of the Galaxy been detected with SPI. This may be regarded as confirmation of earlier expectations that positrons should arise predominantly from sources of nucleosynthesis distributed throughout the plane of the Galaxy, which produce proton-rich unstable isotopes. But there are other plausible sources of positrons, among them pulsars and accreting binaries such as microquasars. SPI results may be interpreted also as hints that these are more significant as positron sources on the Galactic scale than thought before, in the plane and therefore also in the bulge of the Galaxy. This is part of the attempt to understand the surprisingly-bright emission from the central region in the Galaxy, which otherwise also could be interpreted as a first rather direct detection of dark matter annihilations in the Galaxy's gravitational well. INTEGRAL has a unique potential to shed light on the various aspects of positron astrophysics, through its capability for imaging spectroscopy.}

\FullConference{7th INTEGRAL Workshop\\
		 September 8-11 2008\\
		 Copenhagen, Denmark}

\begin{document}

\section{Introduction}
\begin{figure}
\centering
\includegraphics[width=0.48\textwidth]{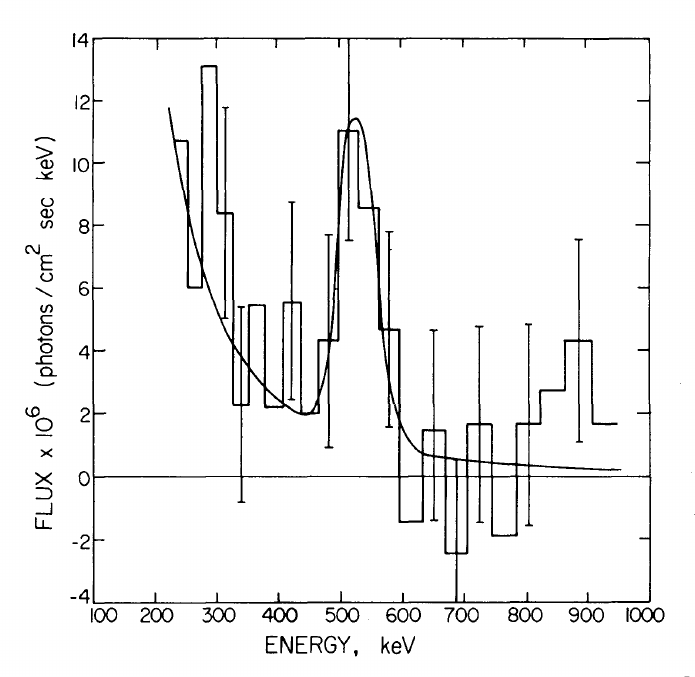} \hfill
\caption{The annihilation line from the inner Galaxy was first detected by a NaI spectrometer in a balloon experiment in the 1970$^{ies}$~\cite{1972ApJ...172L...1J, 1973ApJ...184..103J,1975ApJ...201..593H}
    }
\label{fig_annihilation_spec75}
\end{figure}

The observational study of celestial \gam-ray lines began with Robert Haymes' (Rice University, USA) high-altitude balloon flights of  NaI gamma-ray detectors. Pointing the wide-field instrument in the direction of the galactic center, they detected a spectral-line feature near 500~keV (Fig.~\ref{fig_annihilation_spec75}), which appeared consistent with the electron-positron annihilation line~\cite{1972ApJ...172L...1J, 1973ApJ...184..103J,1975ApJ...201..593H}. Subsequent balloon experiments confirmed this feature, and in particular those instruments with better energy resolution~\cite{1978ApJ...225L..11L} removed any doubt as to the origin of the line. The detection of this characteristic positron annihilation line of cosmic origin had come as a surprise, though in hindsight many potential sources were proposed to contribute detectable numbers of positrons (see below), most plausibly interstellar radioactive decay of freshly-synthesized isotopes ejected by supernova and nova explosions, as well as cosmic-ray interactions in interstellar gas, and pair plasma ejection from pulsars and accreting neutron star of black-hole binaries~\cite{1978PhT....31c..40L}.

Since then, the gamma-ray signature of positron annihilation is being explored. The astrophysical goal is to learn about cosmic acceleration of charged particles and their propagation, which is often associated with e$^{-}$e$^{+}$~pair production, and about other sources of positrons such as nuclear decays. Lately, discussions of dark matter and its annihilation or decay added additional excitement to the study of positron annihilation, as such decay plausibly will also produce e$^-$e$^+$~pairs. Recent measurements of positrons directly in near-Earth space added excitement, from unexpected features seen in the spectral energy distribution of these cosmic-ray positrons. Although only few, specialized gamma-ray telescopes are capable to observe and measure the characteristic gamma-ray signal from cosmic positron annihilation, the uniquely-different underlying physics is the reason that astronomy of positron annihilations now develops into an interesting astrophysical enterprise of its own. INTEGRAL measurements have contributed to re-kindle a broader interest in this topic, in this current epoch of advanced instrumentation and computational tools for this type of research.

In this paper, we review the history of annihilation gamma-ray observations and their interpretation, and present the astrophysical issues of the nature of positron sources and the processes for their annihilation, in the light of the most-detailed observational achievements obtained in recent years with INTEGRAL's SPI Ge spectrometer.

\section{Establishment of the Galactic Positron Annihilation Puzzle}
After the 511~keV line had been established as a signature of the annihilation of positrons from some cosmic origin (see above), numerous balloon-borne instruments were launched, aiming at more details on these cosmic positrons. Apparently-different measured line fluxes were reported, in some cases just months apart \cite{1983AIPC..101..211M}. The common interpretation was that a necessarily compact source (from light-travel time arguments) was episodically ejecting positrons into a relatively dense medium, where they could annihilate quickly, a scenario which was associated with the {\it Great Annihilator}~\cite{1991SciAm.265...29L}. This source had been identified as a prominent and highly-variable radio source, ejecting pc-scale jets~\cite{1992Natur.358..215M}. Gamma-ray variability seemed a plausible consequence of such large-scale plasma jets, and was reported for this source based on SIGMA observations~\cite{1991ApJ...383L..45B,1991ApJ...383L..49S}. It was unclear how positron annihilation would occur, and if nearby molecular-cloud material played a role~\cite{1991A&A...251L..43M}. Other sources of transient annihilation line features were reported as well, such as Nova Musca~\cite{1992ApJ...389L..79G} and 1H~1822-371~\cite{1995ApJ...442..638B}. This supported the notion of variable sources such as accreting binaries as being the most prominent sources of interstellar positrons.

Several years earlier, the Sun had already become the first astrophysical laboratory for the study of positron annihilation~\cite{1976ApJ...210..582C}, after the annihilation gamma-ray line had been detected with a simple NaI instrument aboard the OSO-7 satellite~\cite{1975IAUS...68..341C}. The Solar Maximum Mission, designed for solar flare observations and launched in 1980, featured a Gamma-Ray Spectrometer (GRS) with exceptional stability. Based on detailed measurements with SMM, positrons in solar flares were found to originate from  flare-accelerated particles when hitting the upper photosphere. Nuclear interactions of flare-accelerated protons and ions with atomic nuclei of the photosphere produce radioactive nuclei and pions that decay by emission of positrons, which annihilate locally \cite{1983SoPh...86..395R,2005AGUSMSP21A..09M}.

Studies of candidate positron producers at these times were guided by these results obtained from the Sun, and had predicted significant {\it diffuse} production of positrons within the Galaxy's interstellar medium (e.g.,~\cite{1987ApJ...316..801K}). Positrons from cosmic-ray interactions and radioactive isotopes liberated by nucleosynthesis events were believed to dominate over the contributions from compact stars and the more exotic jet sources. The annihilation process then would depend on the properties of the interstellar gas around the sources~\cite{1979ApJ...228..928B}.
This diffuse distribution was naturally thought to correspond to the Galactic disk, because so many of the most plausible positron sources were located there. These include supernova produced radioactive positron emitters, especially \Al and \Sc (from \Ti) from core collapses and \Co (from \Ni) from thermonuclear events, pairs produced in black hole high temperature accretion disks or jets, pulsar pair winds, and cosmic ray/ISM collision produced positron emitters or $\pi^+$'s, among others.

SMM also detected the positron annihilation line from the interstellar medium of the Galaxy~\cite{1988ApJ...326..717S}. Nearly a decade of tracking the Sun around the sky led to significantly-improved measurements of the positron annihilation line and the associated positronium continuum~\cite{1990ApJ...358L..45S}. The SMM Gamma-Ray Spectrometer instrument measured the largest line flux compared to other experiments designed to trace the activity of the {\it Great Annihilator}. But it was found to be constant over several years~\cite{1990ApJ...358L..45S}. In fact, many of the earlier differing fluxes could be explained by the different fields-of-view of the various instruments viewing a broadly diffuse distribution of annihilation sites~\cite{1991AdSpR..11..203V}. None of these early experiments had imaging capabilities, fields of view ranged from $\sim$15~degrees to $\sim$130~degrees for SMM, but it seemed plausible that the attribution of the measured annihilation gamma-rays to a point source had been in error, and the emission was rather extended.

Thus it was somewhat surprising that early CGRO/OSSE measurements showed that the highest flux measured in its small field came from near the Galactic center \cite{1993AIPC..280..107P}. OSSE had a 11.4-by-3.8~degree collimated field of view, suitable to obtain some spatial information. It was found that the measured annihilation gamma-ray flux from the Galactic center pointings was nearly independent of the direction of the rectangular collimator~\cite{1993ApJ...413L..85P}. The sky distribution of the emission was fit with a compact bulge-like component, and a more extended disk, for which there was only weak evidence then.

As more observations accumulated at a variety of positions and collimator angles, it became possible to derive in a model-independent way distributions of the celestial emission roughly consistent with the data~\cite{1997ApJ...481L..43C,1997ApJ...491..725P}. These showed features similar to the above model: a compact bulge with an intensity profile with FWHM $\simeq 5^\circ$, an extended component along the Milky Way disk, and an excess slightly offset from the galactic center to negative longitude and positive latitude. This third component (sometimes referred to as {\it PLE} for {\it positive-latitude enhancement}) was detected with relatively low ($\simeq$3.5$\sigma$) statistical significance, and much attention followed it~\cite{1997ApJ...487L..57D}. But the origin of the brightest component remained a mystery. More data were acquired, and statistics in both mapping and model fitting improved~\cite{2001ApJ...559..282K,2001AIPC..587...11M}. The significance of the feature at positive latitude diminished. The basic description of the emission remained a bright central bulge and a relatively weak disk.

The line and positronium continuum fluxes were everywhere consistent with annihilation of $\sim$95\% of the positrons via the formation of positronium in the statistical ratio of 3:1 for the triplet to singlet states~\cite{1996A&AS..120C.317K}; the triplet state annihilation continuum was detected with greater significance than the line. (The 3-photon to 2-photon annihilation ratio yields the fraction of annihilations which proceed through the intermediate state of a positronium atom (Ps), the {\it Ps fraction}). Because these remained relative flux measurements due to the background estimation from offset pointings, the absolute fluxes of bulge and disk components were not tightly constrained. The central component could be compact, like the stellar bulge, with a relatively low total flux, or more extended, like the halo stellar surface brightness, with a larger total flux. Similarly, a very thin disk with a low flux fit nearly equally as well as a vertically extended disk with much larger total flux \cite{2004NewAR..48...69M}. The ratio of bulge flux to disk flux, and by assumption, positron production rates, were found to be of order unity, with an uncertainty almost a factor of three. The flux of each component was about $7\ 10^{-4}$ \cms, with a systematic uncertainly in each of roughly a factor of two. Note that both components cannot be pushed to their lower limit together, which would be inconsistent with the wide-field measurements, that of SMM/GRS in particular.

Any such bright bulge was unexpected, as most of the prominent sources are in the young stellar population of the disk (see below). Even older population sources, such as Type Ia supernovae, would be expected to inject substantially more positrons into the massive disk than the bulge. It seemed that multiple sources are required, and known or very likely sources such as the decay of \Al and \Ti should be subtracted from the disk, raising the bulge to disk flux ratio for the unknown source(s). The annihilation line flux from \Al positrons is F$_{511}$ =  0.47~F$_{1809}$ for the above positronium fraction, so the observed 1809 keV flux~\cite{1995A&A...298..445D} indicated that 0.1 to 0.5 of the disk annihilation flux would be due to \Al decay. The Galaxy-wide contributions from the \Ti decay chain are not so directly observed, but various arguments (outlined below)  suggested that the average rate of \Ti production provides a line flux at 511 keV of $2\ 10^{-4}$ \cms, coincidentally similar to that from \Al. Still it seemed that an additional source of positrons may be required in the disk, and a still more productive source in the central bulge or halo. Puzzling indeed.

\section{Positron Sources and Annihilation Gamma-Rays}
There are considerable uncertainties in estimates of the positron yields from all candidate sources that have been mentioned above. Additionally, positrons may propagate in interstellar space away from their sources, before they annihilate. For these reasons, discussions about potential origins of the annihilation gamma-rays are centered on rather global properties, such as variations in time and space and in particular the relative brightness of the Galaxy's bulge and disk. Here we summarize the main characteristics of positron emission from the sources as much as we know them (see also Table~\ref{tab_candidate_sources}).

\subsection{Positron Sources}
{\bf Cosmic-ray nuclear reactions} will produce positrons as an intrinsically-diffuse source throughout interstellar space, following the morphology of gas density throughout the Galaxy. The main process is the decay of $\pi^+$, which arise from hadronic $p+p\rightarrow\pi^++X$ reactions. Positron energies will reflect the energies of those secondary pions, hence be distributed around an energy of $\sim$100~MeV. The total Galactic positron production from this source is estimated as 1--2~10$^{42}$~e$^+$s$^{-1}$. This source potentially can be independently verified through direct cosmic-ray measurements in near-Earth space.

{\bf Pulsars} will emit electron-positron pairs along their open magnetic-field lines and in a {\it pulsar wind}. Positron energies depend on the nature of the accelerating potential {\it gap} which develops from the rotation-induced loss of charges at the outer magnetosphere -- {\it polar} and {\it outer gap} models are being discussed \cite{1983ApJ...266..215A,1973Natur.246...93R}. Charged-particle acceleration in these gaps leads to electromagnetic cascades and {\it curvature-radiation} photons, as the accelerated charged particles are forced on curved trajectories by the strong magnetic field~\cite{1982ApJ...252..337D}. Since most of these photons are well above the pair-creation threshold energy, secondary electron-positron pairs are produced as photons propagate outward in the pulsar magnetosphere and collide with each other and/or with X-ray photons emitted from the hot neutron-star surface. Power-law type photon spectra reaching up to and beyond 10$^{15}$eV are plausible and supported by gamma-ray observations from pulsars~\cite{2005AIPC..745..105H}. Positrons may escape through open field lines at the outer magnetosphere, or through the pulsar wind. Young pulsars with high magnetic fields are expected to eject 10$^{36}$--10$^{40}$~e$^+$s$^{-1}$ during their first $\sim$1000 years, most of which should annihilate in the surrounding nebula. If the $\sim$2000 pulsars known mostly from radio observations~\cite{2005AJ....129.1993M} have ages up to $\sim$10$^5$ years and comprise $\sim$1\% of the true pulsar population, this may result in a positron production of up to $\sim$~10$^{43}$~e$^+$s$^{-1}$. On the other hand, in the nebular phase, positrons will be confined, and escape into the surrounding ISM is expected to occur for ages above 10$^5$~y, after some fraction annihilated within the nebula. The production of these (older) pulsars will be significantly lower, as their magnetic field is weaker. The recently-detected excess in high-energy positron flux near Earth with the HEAT, PAMELA, and ATIC instruments has been attributed to the production of nearby pulsars, and may provide additional constraints on their total production~\cite{2008arXiv0812.4457P,2009arXiv0903.1310M}. Pulsar emission can be studied in detail throughout the electromagnetic spectrum from radio to high-energy gamma-rays, and is being exploited in order to build a physical model of the pulsar magnetosphere with its charge currents and its particle acceleration; the general picture of an electromagnetic cascade developing from curvature radiation of accelerated particles and leading to a pulsar wind appears rather solid, yet insufficient in detail for quantitative estimation of the positron source characteristics. Detailed measurements of the space distribution of pulsars are available, e.g. the ATNF catalogue~\cite{2005AJ....129.1993M}. Evident is that the pulsar population is extended in latitude well above the gas disk of the Galaxy. This is consistent with the pulsar velocity distribution, which shows that pulsars apparently receive a significant {\it kick} at birth. Unclear remains the bias in such a catalogue of radio pulsars with respect to the positron-producing variety, i.e. the young pulsars (a<10$^3$y), or the {\bf millisecond pulsars} spun up at late times due to accretion from a binary companion, or {\bf neutron-star binary systems (LMXB, HMXB)} with suitable mass transfer properties so they may form a jet or another kind of plasma ejection. Although the X-ray emission of such binaries has been studied in great detail and led to general understandings of the evolution of binaries and transient accretion-flow changes, an ejection of positrons into interstellar space is rather unclear, especially when no plasma jet is seen (see e.g. \cite{2006ChJAS...6b.268W} for millisecond pulsars). Another main uncertainty on all neutron-star-related positron sources arises from the neutron star birth rate, with an estimate of 1/1000~y for pulsars, and $\sim$1/50~y as the best current estimate of the rate of core-collapse supernovae producing a compact remnant star in our Galaxy \cite{2006Natur.439...45D}.

{\bf Microquasars} would constitute another candidate source of positrons. These objects are characterized by a plasma jet in which superluminal motion can be traced in radio and IR emission and appears linked to earlier X-ray events characterizing accretion-flow irregularities near the compact object~\cite{1992Natur.358..215M, 1994Natur.371...46M, 2004MNRAS.355.1105F}.  
Although the plasma-jet composition is unclear, observed synchrotron emission demonstrates the presence of electrons (and/or positrons), and associated energies suggest that pair production could plausibly occur in this environment, both in the jet base and in jet interactions with surrounding interstellar gas. Therefore, {\it microquasars} are considered plausible positron sources. Positron energies are correspondingly uncertain, ranging from mildly-relativistic MeV energies to the ultra-relativistic energies of the jet plasma. Per microquasar, a positron flux on the order of 10$^{41}$ e$^+$s$^{-1}$ has been estimated~\cite{2005A&A...436..171G}. A few {\it microquasars} have been clearly-identified within our Galaxy, most notably 1E1740.7-2942, GRS~1915, SS433, and LS~5039. A small but significant fraction of the ultra-luminous X-ray sources seen in other galaxies in abundance (up to hundreds) may also be microquasars with relativistically-beamed jets. Within the Galaxy, hence, the total positron annihilation luminosity of $\sim$10$^{43}$s$^{-1}$ could well be provided by microquasars~\cite{2005A&A...436..171G}.

{\bf The Galaxy's Supermassive Black Hole (Sgr~A$^*$)} constitutes an extreme variant of similar type, and a candidate source of positrons~\cite{2006ApJ...645.1138C, DogielPoS}. Mass accretion onto this supermassive black hole could lead to formations of either a hadronic (proton-dominated) or leptonic (electron-positron-plasma dominated) jet. For each of these hypotheses, models and evidences have been discussed~\cite{2006ApJ...645.1138C,2007PThPS.169..117T}. So, a proton-dominated jet could exist largely unseen, and produce positrons through secondary pions as the proton jet impacts on a nearby molecular cloud \cite{2006ApJ...645.1138C}. Alternatively, plasma ejection from past more-intense accretion episodes is plausible, e.g. from the brightness of reflection nebulae, and the expanding {\it `molecular ring}~\cite{2007PThPS.169..117T}. The current accretion (and hence positron ejection) activity is probably weak, from the low luminosity at the position of Sgr~A$^*$~\cite{2006ApJ...636..275B}. The immediate surroundings of Sgr~A$^*$ is rather devoid of gas; this may explain the currently-low activity, and be consistent with absence of bright sources of high-energy emission. But rather dense interstellar clouds at $\sim$few hundred pc distance should stop and annihilate positrons ejected from the Sgr~A$^*$ surroundings, and it is not obvious how the annihilation gamma-ray source could be as symmetric and extended as observed (see however \cite{ChernyshovPoS,DogielPoS}).

{\bf Nucleosynthesis events} will generally produce new atomic nuclei, some of those being created on the proton-rich side of the stable isotopes, so that their radioactive decay ($\beta^+$-decay) liberates positrons. Typical positron energies from radioactive decays are $\sim$MeV. Most-plausible origins of proton-rich isotopes are hydrogen-burning nucleosynthesis events such as novae, with $^{18}$F and $^{13}$N as two prominent isotope products~\cite{1999ApJ...526L..97H}. But also nuclear burning of rather symmetric matter (i.e., equal numbers of protons and neutrons) up to the iron-peak region inside supernovae will lead to positron emission from $\beta$-decays, e.g. from $^{44}$Ti produced inside core-collapse supernovae and $^{56}$Ni produced in all types of supernovae. A key uncertainty arises from the originally-dense production sites for the unstable isotopes in these cases; depending on the characteristics of overlying stellar envelope material and on the structure of magnetic fields, the positrons from radioactive decays could annihilate locally within the sources, rather than escaping into the interstellar medium; then, annihilation gamma-rays could only be visible if the overlying envelope would be optically thin to these gamma-rays. For example, escape rates for SNIa are typically a few percent only~\cite{1999ApJS..124..503M,1993ApJ...405..614C,2007ASPC..372..407H}. Nevertheless, the summed contributions of these positron sources has been considered in agreement with the steady-state annihilation rate as observed in the 511~keV gamma-ray line, i.e. 10$^{43}$e$^+$s$^{-1}$~\cite{1992ApJ...397..135S, 2002NewAR..46..553M, 2006A&A...449..869P}. The spatial distributions of sources within the Galaxy should follow the stellar disk, although characteristic deviations in detail are expected from the nature of the nucleosynthesis events: Supernovae from massive stars will correlate with the young stellar population, their spatial distribution in the current-day Galaxy is traced best by the gamma-ray map of $^{26}$Al gamma-rays from these same sources~\cite{2001ESASP.459...55P,1995A&A...298..445D}, with a dominant galactic-ridge contribution ranging in Galactic longitudes from at least -60$^\circ$ (Carina region) to +80$^\circ$ (Cygnus).
The total positron yield of classical novae is small~\cite{1988ApJ...328..755L}; they could have a more prominent bulge contribution~\cite{2006MNRAS.369..257D}, as also
does the common model for the spatial distribution of SNIa, although both will also have significant ($\sim$50\%) populations within the disk of the Galaxy (Note that the SNIa population is modeled to arise from a {\it prompt} component related to the star formation activity directly similar to massive stars, and a {\it tardy} component which is delayed by 3--4~Gy and resembles the duration it may take to increase a white dwarf's mass by slow accretion from the companion star to the Chandrasekhar limit~\cite{2006MNRAS.370..773M}). Therefore, if positrons would annihilate near their sources, this appears incompatible with the bulge-dominated image from annihilation gamma-rays.

\begin{figure}
\centering
\includegraphics[width=0.48\textwidth]{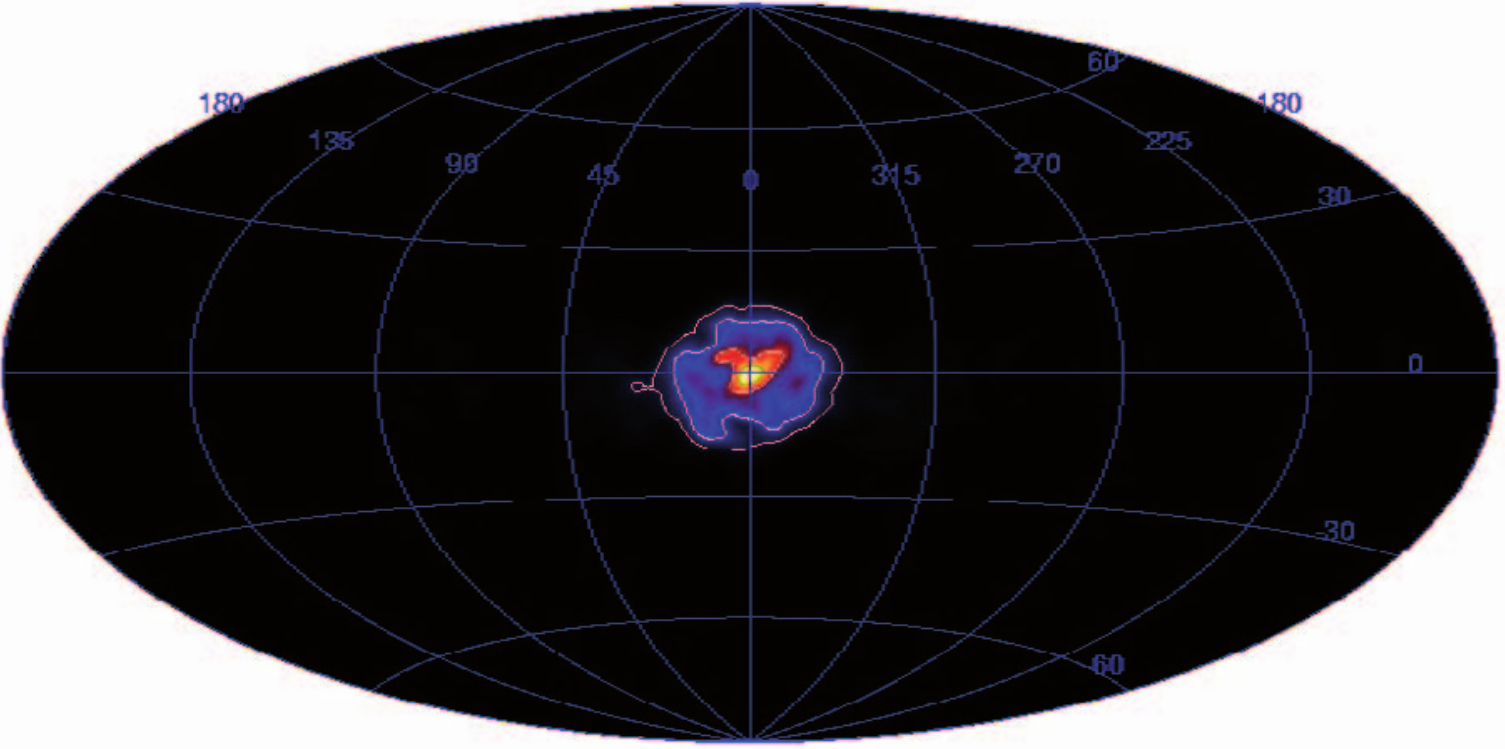} 
\includegraphics[width=0.48\textwidth]{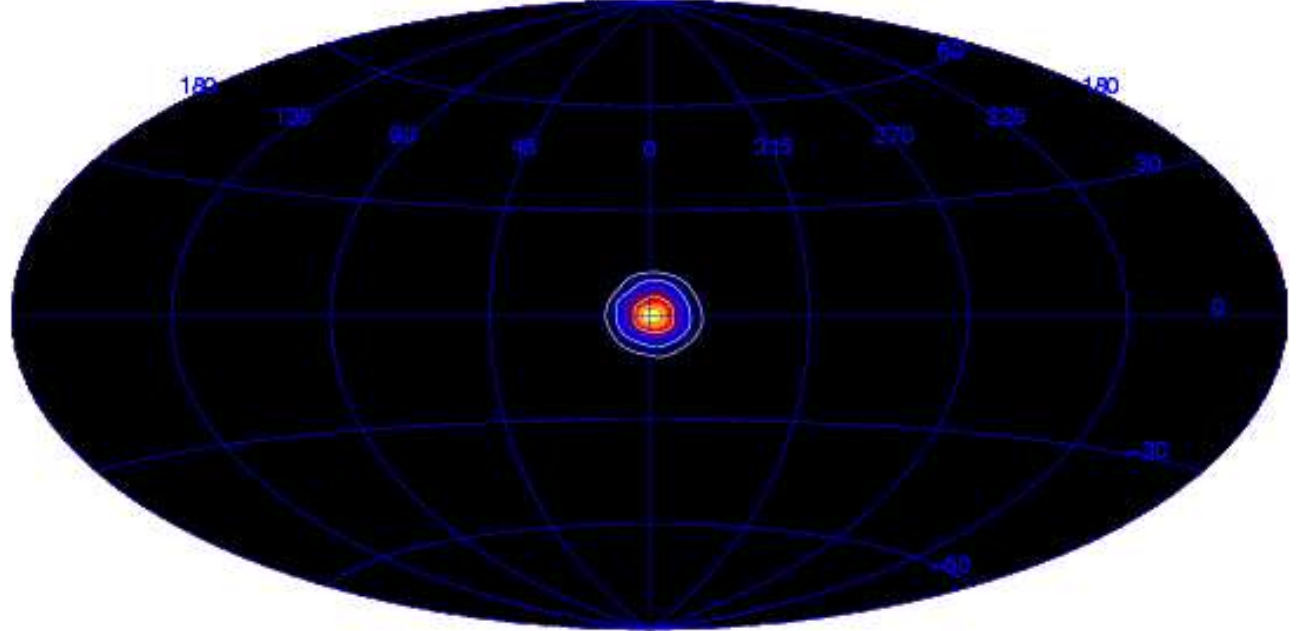}\hfill
\includegraphics[width=0.48\textwidth]{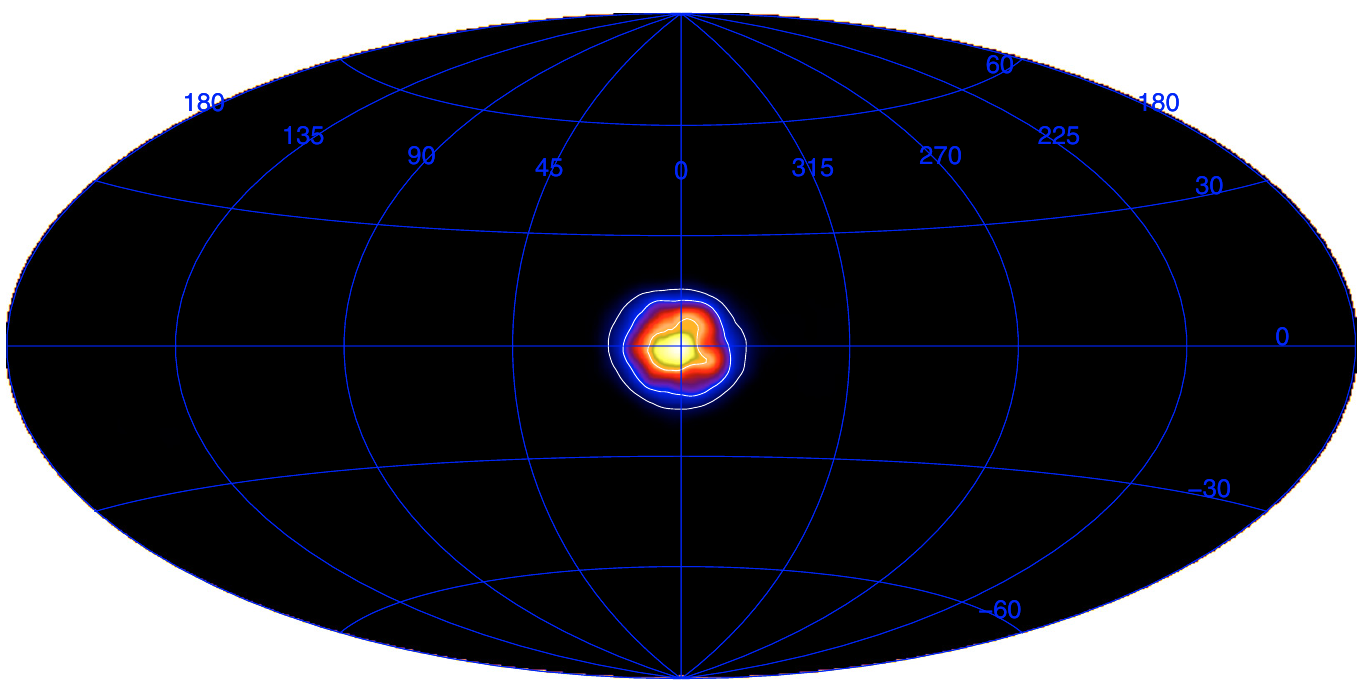} 
\includegraphics[width=0.48\textwidth]{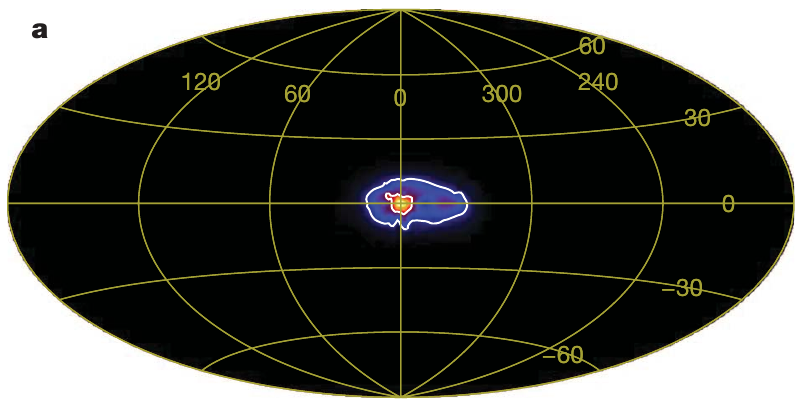}
\caption{Positron annihilation maps from INTEGRAL/SPI observations:
      The RL image from one year of data in the 511~keV line~\cite{2005A&A...441..513K} ({\it Top Left}).
    ({\it Top Right:}) The MREM image from almost three years of data~\cite{2007astro.ph..2621W} ({\it Top Right}).
      The RL image from one year of data in Ps continuum emission~\cite{2006A&A...450.1013W} ({\it Bottom Left}).
    ({\it Top Right:}) The RL image from almost five years of data  in the 511~keV line~\cite{2008Natur.451..159W} ({\it Bottom Right}).
    }
\label{fig_annihilation_maps}
\end{figure}

{\bf Positrons from Dark-Matter decays} have been discussed as a possibility, mainly because a rather spheroidal, symmetric, bulge-centered annihilation emission morphology (as observed indeed; Fig.~\ref{fig_annihilation_maps}) would be characteristic for such a positron origin~\cite{2006MNRAS.368.1695A}. Since dark-matter particles by definition would interact weakly and hence be collected by the global gravitational potential of our Galaxy in large abundance for a small actual annihilation rate, the nature of dark-matter particles would not be too important, nor would be details of the (small) deviations from symmetry in the Galaxy's gravitational potential. As the annihilation rate is proportional to the square of the dark matter particle density, the central part of the gravitational potential (i.e., the inner kpc of our Galaxy) is most important for possibly observable signatures; this is the region where theoretical uncertainties are being discussed, e.g. deviations from the proposed {\it Navarro-Frenk-White (NFW) potential} form~\cite{1997ApJ...490..493N} and a possible flattening of the power-law profile of the spherical halo within the inner $\sim$100~pc.  Different particle candidates are being discussed. From theoretical motivations, weakly-interacting supersymmetry particle candidates in the GeV--TeV mass range are preferred; the minimalistic supersymmetry extension of the Standard Model combined with constraints from high-energy particle physics experiments and the large-scale structure of the universe currently favor a lightest-dark matter spin-1/2 particle as dark-matter candidate with a mass in the 30--600~GeV range~\cite{1984NuPhB.238..453E}, the lightest such {\it neutralinos} could decay with positron production as minor by-products~\cite{2004ESASP.552...65C}, e.g. through co-annihilation modes; but constraints from unobserved but stronger high-energy gamma-ray signatures of such neutralino decay rather favor light scalar particles as an attractive possibility, as they could solve some inconsistencies in the standard model without invoking supersymmetry~\cite{2004NuPhB.683..219B}. Their mass in the few-MeV range would be consistent with present observational constraints, as these avoid the production of continuum gamma-rays from Bremsstrahlung and annihilation-in-flight which is tightly constrained by gamma-ray observations~\cite{2006PhRvL..97g1102B}. Since all of this is {\it new physics}, the annihilation modes of dark matter are open to adjustments of the theory of dark-matter particle nature and properties, with few global constraints only. As a result, positron energies and annihilation cross section are largely unknown, and the spatial symmetry of a signal in the Galaxy is the strongest motivation for discussing dark-matter origins of cosmic positron annihilation gamma-rays. Nevertheless, if other positron annihilation channels and sources can be sufficiently constrained or excluded, positron annihilation gamma-rays would constitute a first clear and rather straightforward observation of dark matter in the Galaxy.

\begin{table}
  \centering
  \caption{Candidate sources of positrons and their main characteristics in terms of positron intensity, positron energies, and spatial distribution in the Galaxy. {\small (PL= powerlaw spectrum; GC=Galactic Center)}}\label{tab_candidate_sources}
\begin{tabular}{lllll}
  \hline
  candidate               & e$^+$ production & e$^+$ energies & locations & comment \\
                          & [e$^+$s$^{-1}$] &          &           &  \\
  \hline
  cosmic-ray interactions & 1--2~10$^{42}$ & GeV  & gas disk  & PL $\alpha\sim$2 \\
  pulsars                 & 10$^{39}$--10$^{43}$ & GeV...TeV & young-star disk & PL $\alpha\sim$1.5 \\
  microquasars            & few 10$^{41}$ &  GeV...TeV & stellar disk and bulge & \\
  SgrA                    & few 10$^{42}$ &  GeV...TeV & GC & when active \\
  nucleosynthesis         & 10$^{43}$ & $\leq$MeV & young-star disk & \\
  dark-matter decay       & ? & $\sim$TeV & spheroid / bulge & \\
  \hline
\end{tabular}
\end{table}

\subsection{From Positron Sources to Annihilation Gamma-Rays}
Maps of annihilation gamma-rays are often naively misinterpreted as being maps of positron sources. The pathways of positrons in interstellar medium around their sources may be complex, and the spatial correlation of annihilation emission and positron production may be close for some, and nonexistent for other positron sources.

As discussed above, stellar sources of positrons such as novae and supernovae will release positrons within their expanding envelopes. Depending on the density structure within these envelopes, and for more diluted late explosions also on the morphology and intensity of magnetic fields, positrons could either interact with envelope material to be slowed down and annihilate, or else escape into the surrounding interstellar medium. From pulsars and jet sources, positrons could be dispersed as a relativistic-particle stream, or in a wind; in the first case, cosmic-ray transport physics applies, while in the wind case, adiabatic expansion energy losses also are to be accounted for. The release of pair plasma into the surrounding medium will generally be characterized by a rather diluted environment, which has been swept up by previous stellar activity (the jet itself, or the supernova explosion preceding pulsar formation). The structure and morphology of this surrounding medium then is the key to near-source annihilation or long-distance propagation, as the two extremes of positron fates.

On parsec scales or larger, the path lengths of escaping positrons depends critically on morphologies of gas and magnetic fields (due to energy losses from ionization and Coulomb scattering and spiraling around magnetic field lines with pitch-angle scattering at magnetic-field irregularities), and on positron energy (from the same processes, but also from resonances with plasma waves).
Positron transport and annihilation in interstellar medium has been estimated in theoretical studies~\cite{1991ApJ...378..170G,2007astro.ph..2158G}. For example, Gillard et al.~\cite{2007astro.ph..2158G} have re-evaluated trajectories of positrons, and find that, for MeV positron energies, path lengths of survival before annihilation can be as large as a few kpc in regular magnetic fields, limited by pitch angle scatterings and Coulomb collisions with ISM gas (here assumed to have density 1~cm$^{-3}$; correspondingly longer for more diluted regions). This same study also shows the crucial role of positron energy and magnetic-field interactions: There is a resonance region where path lengths can be large, but for most other conditions (positrons at high energy with ionizing and Coulomb collisions; molecular-gas clouds along the trajectories) positrons will not propagate beyond 50--100~pc, and mostly annihilate in the warm outer regions of molecular clouds. This leaves us with a picture whereby local annihilation may be a reasonable assumption for the degree-type precision of gamma-ray maps, unless the positron production region has been pre-shaped into a {\it champagne-flow} from previous activity shaping the ISM around the sources, which would lead to substantial positron escape into the Galaxy's halo.

The (uncertain) morphology of the halo's magnetic field~\cite{2008NuPhS.175...62H} then may have an interesting effect: Positrons could be channeled by a large-scale symmetric field of dipole structure to enter the denser regions of the Galaxy's ISM preferentially in the central part of the bulge~\cite{2006A&A...449..869P}. Although this is an attractive speculation to explain the relatively bright annihilation emission from the bulge/inner Galaxy, it is unclear if the magnetic field could be as symmetric and ordered as assumed here; radio arcs do suggest magnetic field structures perpendicular to the plane of the Galaxy in its central region, but on the other hand the gas distribution is far from spherically-symmetric, and should be reflected in corresponding spatial structure of the annihilation emission.

\section{INTEGRAL Results and Recent Developments}
With the INTEGRAL mission, a new level of observational capabilities was introduced to the study of Galactic positron annihilation.
Its relatively large field-of-view accumulates exposure to the diffuse emission efficiently, especially given the INTEGRAL mission's emphasis on the inner Galaxy.  The $\sim$3~degree spatial resolution of SPI with its coded mask allows to generate a map of annihilation gamma-ray emission. 
The 2~keV energy resolution of SPI's Ge detectors (at an energy $\sim$500~keV) is a major improvement over previous instruments with comparable sensitivity, and allows to set additional constraints, as the environmental conditions of positron annihilation leave their imprints on the 511~keV line shape and the line-to-continuum intensity ratio.

INTEGRAL/SPI first rediscovered and confirmed many of the above aspects of the electron-positron annihilation radiation from the Galaxy. The gamma-ray intensity map generated in the 511~keV line from 15~Msec of data of the first INTEGRAL mission year showed the remarkably-bright and extended central emission region to be rather symmetric and centered in the Galaxy, with an extent both in longitude and latitude of about 8$^o$~(FWHM) and a flux of (~$10^{-3}$~\cms)~\cite{2005A&A...441..513K}. No asymmetries such as the earlier-reported 'positive-latitude enhancement' was seen, nor any remarkable emission from the Galaxy's disk itself. This only showed up significantly in model fits and when more data were integrated~\cite{2007astro.ph..2621W}, as the surface brightness is low, and instrumental background rather high (for bright regions, the celestial 511~keV signal reaches 2\% of the total count rate only~\cite{2005A&A...441..513K}). The overall brightness of the disk is constrained to lie in the range (0.9--1.8)~10$^{-3}$~\cms, which corresponds to bulge-to-disk flux ratios in the range 0.4--1.4~\cite{2007astro.ph..2621W}. The flux of the bulge-like emission was found similar to the CGRO/OSSE value~\cite{2005A&A...441..513K}, with emission being of a slightly larger extent, 8$^{\circ}$ FWHM as derived from a fit with a two-dimensional angular Gaussian. Thus SPI data confirm the diffuse, rather than point-source dominated, emission, and the inner (bulge) region being much brighter than any emission from the Galactic disk.

\begin{figure}
\centering
\includegraphics[width=0.7\textwidth]{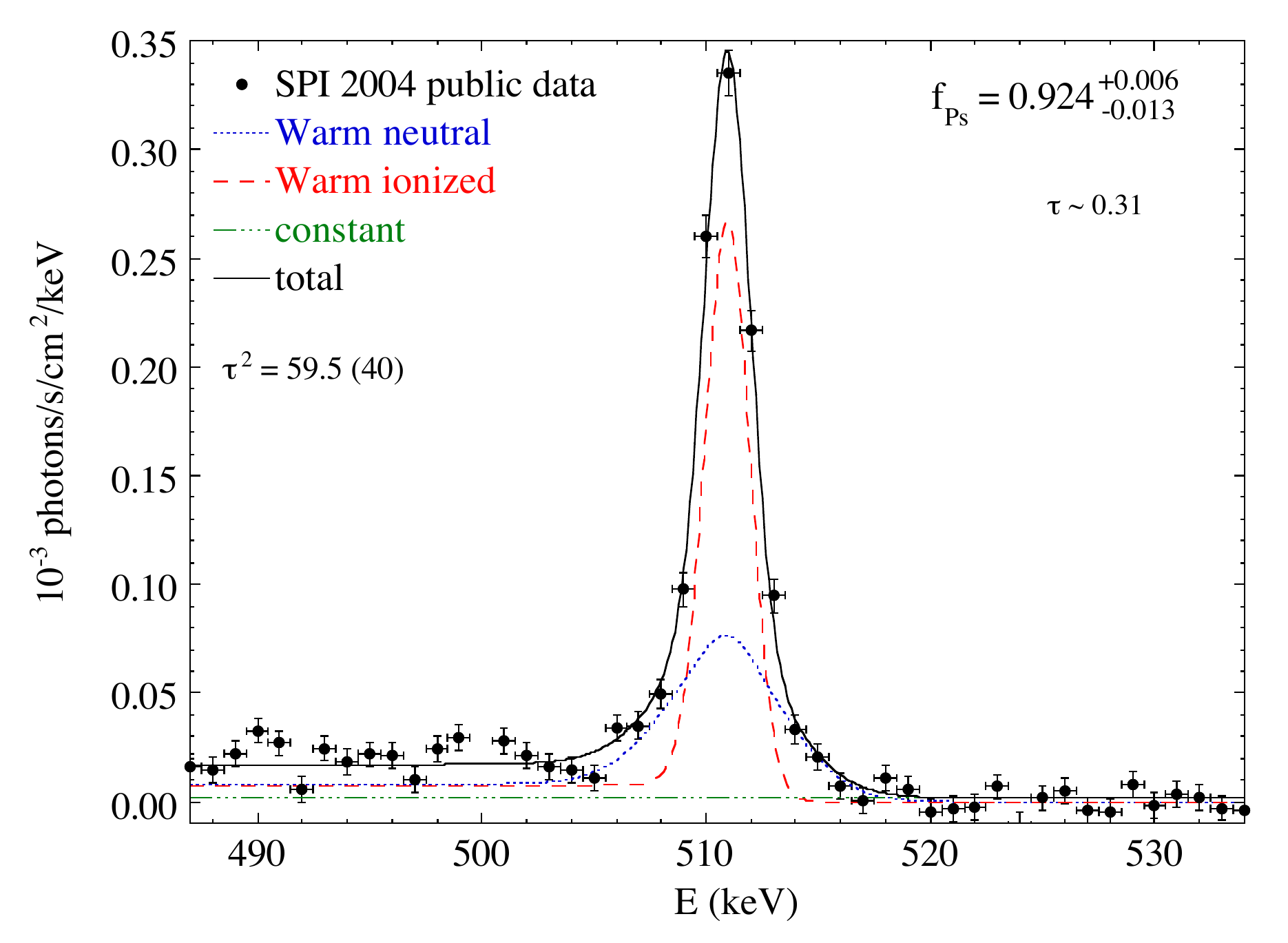} \hfill
\caption{The annihilation line from the inner Galaxy as decomposed into different components related to phases of the ISM~\cite{2004ESASP.552...51J,2006A&A...445..579J}}
\label{fig_annihilation_spec_SPI}
\end{figure}

It is reassuring that imaging analysis of the positronium continuum confirms this overall morphology, although in detail differences have been noticed, which appear consistent with the spectral energy distribution differences between bulge and disk regions (see below)~\cite{2006A&A...450.1013W}.  The 3-photon to 2-photon annihilation ratio yields the fraction of annihilations which proceed through the intermediate state of a positronium atom (Ps). This Ps fraction has been re-determined from the improved spectral precision of INTEGRAL/SPI's Ge detector measurements, and is confirmed to lie beyond 90\% and up to 100\% ((96.7 $\pm$2.2) \%~\cite{2006A&A...445..579J,2005MNRAS.357.1377C,2007astro.ph..2621W,2008ApJ...679.1315B}). This implies that partially-neutral gas is a characteristic of the medium where positrons annihilate.

More precise constraints can be derived when the imprints on the shape of the 511~keV line is also included in an analysis to determine the ionization and temperature of the interstellar gas at the annihilation sites. At the extremes, the hot phase of the ISM would incur line broadening to a level of $\sim$~10~keV (not observed), while annihilations on dust would suppress any such thermal Doppler broadening through momentum transfer to a heavy dust grain~\cite{2004ESASP.552...57G}. Parametrizing the ISM globally with these two parameters (ionized fraction and temperature), \cite{2005MNRAS.357.1377C} find a best match to the SPI spectrum for temperatures near 8000~K and 10\% ionization. This excludes the hot parts of the ISM with temperatures well in excess of 10$^4$K as well as molecular clouds, while outer cloud fringes appear a plausible annihilation site. Of course, a composite of different annihilation sites also is plausible, given the transient and complex morphology of the interstellar medium~\cite{2005A&A...436..585D}. 
Jean et al. 
\cite{2006A&A...445..579J} fit a multi-component ISM to the measured spectral shape (Fig.~\ref{fig_annihilation_spec_SPI}), and find that the {\it warm neutral} and {\it warm-ionized} phases hold $\sim$50\% shares, while {\it cold} and {\it hot} ISM phases contributions are minor; this may be the result of volume weighting, as filling factors of the various ISM phases are of such magnitudes. It is not obvious how these findings can be related to the locations of positron annihilation with respect to their sources, in particular since recent investigations of the ISM cast doubts on the definition of such {\it phases}~\cite{2005A&A...436..585D,2007ApJ...665L..35D}.

\begin{figure}
\centering
\includegraphics[width=0.99\textwidth]{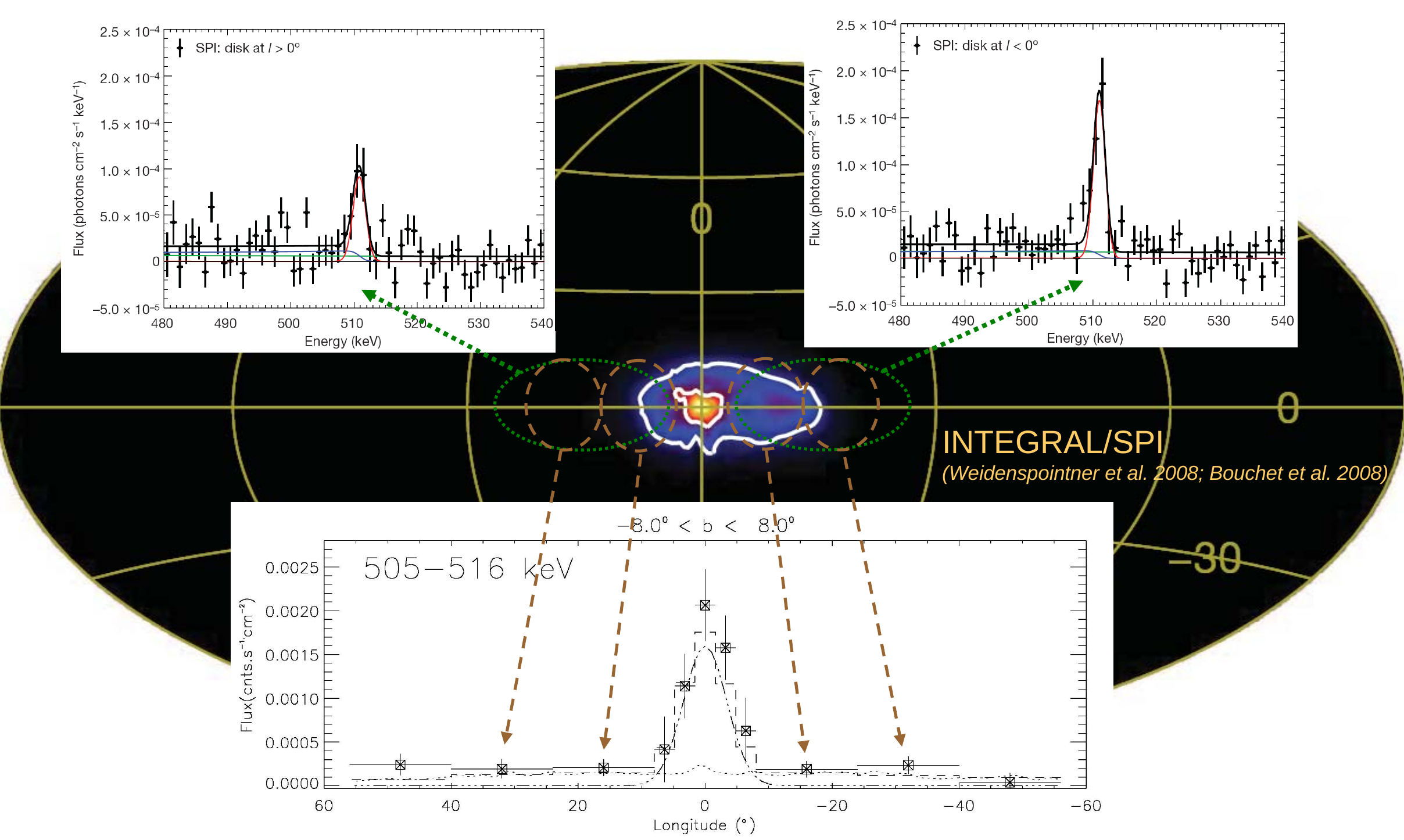} \hfill
\caption{The annihilation line from the disk of the Galaxy is now seen clearly, in addition to the bright bulge region. An asymmetry favoring the inner-disk part in the fourth Galactic quadrant is consistently found, depending on the approach of instrumental-background modeling very clearly in one analysis~\cite{2008Natur.451..159W} ({\it top}), and as a hint only in an alternative analysis~\cite{2008ApJ...679.1315B} ({\it bottom})}
\label{fig_annihilation_asymmetry}
\end{figure}

Analysis of more than three times as much SPI data (5.4$\ 10^7$ s) have been performed now~\cite{2008Natur.451..159W,SkinnerPoS}. Above results are confirmed, with similar bulge fluxes (~$10^{-3}$ \cms) and extent (6$^\circ$ FWHM). Yet, since observational strategy and analysis method are the same, these results are subject to the same systematic uncertainties as 
before. In particular, in the early mission, pointings were not sufficiently randomized, so that
background 
trends in time could correlate with trends in Galactic longitude, for example. Therefore, a different observational-pointing strategy of rapid swings through the Galactic plane at steps in longitude has been introduced for more recent observations since 2007; these should provide a useful test for systematics effects, and comprise a systematically-different view at the Galactic-plane distribution of annihilation emission. Nevertheless, from 4.3 years of data, the disk is now clearly detected, and evidence is presented that the disk is asymmetric, with the negative longitude inner disk brighter than the positive longitude side~\cite{2008Natur.451..159W}. The inner disk is fit with a broader (7$^\circ$) latitude profile, so the total flux is comparable to that of the bulge, and the negative longitude side is brighter by a factor 1.8. An independent analysis of nearly the same data set, but using a different algorithm to derive the background estimate, finds a quite similar bulge extent and flux~\cite{2008ApJ...679.1315B}; the disk asymmetry is only hinted at in this work, while consistent. The disk receives a large flux of 1.7$\ 10^{-3}$ \cms in the latter fit to observations, and
has 
a latitude extent 15$^{\circ}$--20$^{\circ}$ at this larger scale (compared to the {\it inner} disk above). On the other hand, the asymmetry result in \cite{2008Natur.451..159W} is confirmed when more data are added (Skinner et al., this conference).  It remains true for these SPI data that the measured fluxes are model dependent for each component. Lower flux is obtained for a more compact bulge, and similarly, the latitude extent of the disk emission is poorly constrained; wider disks have significantly higher fluxes.
Clearly we are still learning how to best analyze these data, and substantially more data will be obtained with an alternate observing strategy, so we can expect further clarification and possible surprises.

\section{Open Issues and Prospects}
Progress is slow in understanding the origin of the positrons which shine so brightly in the inner part of our Galaxy through their annihilation gamma-rays. How many are due to nuclear processes (nucleosynthesis or cosmic-ray origins)? How many have their origins in pair plasma (pulsars and/or accreting binaries/SgrA)? The amazing symmetry of the bright inner emission region had stimulated much excitement in dark-matter searches, because it seems difficult to obtain such extended emission (with similar longitudinal and latitudinal extents) yet well-centered in the Galaxy from stellar sources. Some caution is appropriate, because imaging algorithms for such background-limited data are often regularized for smoothness in order to stabilize the iterative image extractions (see Fig.\ref{fig_annihilation_maps}). Nevertheless, this dominant characteristic of Galactic annihilation emission needs to be explained: why does positron annihilation occur predominantly in the bulge region of our Galaxy?

\begin{figure}
\centering
\includegraphics[width=0.8\textwidth]{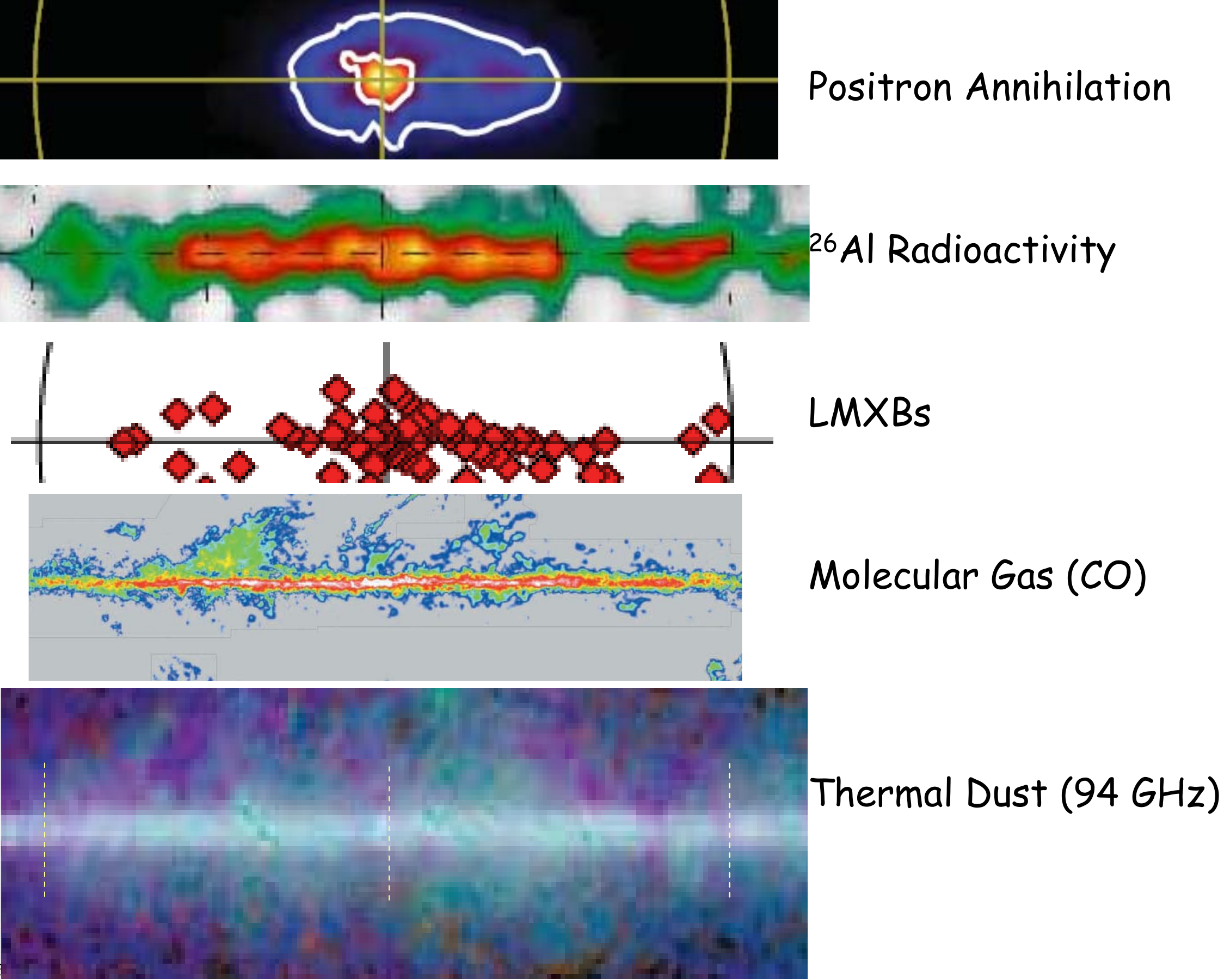} \hfill
\caption{Comparisons of the morphology of positron annihilation emission along the plane of the Galaxy with tracers of candidate positron sources may help to illuminate their different contributions within the disk of the Galaxy, and by inference, any residual annihilation in the bulge which might need more exotic explanations such as large-scale positron transport through the Galaxy's halo, or annihilation of dark-matter particles.}
\label{fig_annihilation_proxies}
\end{figure}

One approach to answer this question is to determine the annihilation emission from known positron sources (
Fig.~\ref{fig_annihilation_proxies}). 
We have mapped \Al emission and know its associated positron production. We note that the \Al longitude profile shows an asymmetry in longitude in the same sense as the annihilation line flux, though much less pronounced. If we adopt \Al as a proxy for massive stars in our Galaxy, core-collapse-produced \Ti decay 
positrons should follow the same spatial distribution. Which other sources are important contributors in the disk of the Galaxy? Pulsars should be, at least on rather undisputed theoretical grounds, and they would probably follow the same large-scale
longitude 
distribution, deriving from core-collapse supernovae. Binary systems are candidates, although this has not been established as firmly. Their spatial distribution could be different. One may take hard-spectrum sources among low-mass X-ray binaries as a proxy, and indeed interesting similarities have been reported concerning Galactic-disk asymmetries~\cite{2008Natur.451..159W}. These sources are also found in the galactic bulge, so they could possibly explain that positron component, but it seems that the ratio of bulge positrons to disk positrons, after subtraction of the \Al and \Ti contributions, is still larger than the ratio of numbers of hard LMXB sources. However, even if this correlation proves to be true, we do not know whether these hard sources eject the positrons themselves, or whether they might be tracers of another population that does.  Another plausible source of positrons in the old bulge is \Co produced in Type Ia supernovae. Only a few percent of the positrons produced at the expected rate need to escape the ejecta to contribute significantly to the observed signal. This is not implausible, but the expected rate of SN~Ia in the disk is thought to be significantly larger than that of the bulge, so if the SNe are not intrinsically different, the relative positron annihilation rates in the two components rule out these as the dominant source.

We may have to search for annihilation emission from near each of these candidate Galactic source objects, in order to estimate their contributions and thus establish their large-scale annihilation brightness, adopting a spatial-distribution model. Cosmic-ray transport models might be important, if local annihilation turns out to be rather small. Alternatively, annihilation emission from localized regions where candidate sources are better understood and different from the inner Galaxy, such as the Cygnus region, could provide new insights. Can we detect enhancements of annihilation gamma-ray emission from spiral arms? Finally, when known sources are well-constrained so that their contributions can be subtracted and we would still be left with a bright annihilation gamma-rays from the bulge, one of the few bulge-only sources, low-mass dark matter particles that decay or annihilate into pairs~\cite{2004NuPhB.683..219B} is an interesting possibility, but not yet constrained by the \gam-ray data.

The gamma-ray sky appearance is very different in annihilation $\gamma$-rays compared to adjacent and most other parts of the electromagnetic spectrum. It may provide a unique opportunity to learn about new astrophysical processes, new types of sources, and at least about the propagation of leptons of relatively-low energies in the Galaxy's interstellar medium.

{\bf Acknowledgements}
This review was stimulated from two topical workshops held at ISSI in Bern (CH), organized by Nikos Prantzos and R.D. in 2007/2008. R.D. acknowledges discussions with many colleagues on this exciting subject, notably Celine Boehm, Laurent Bouchet, Andrei Bykov, Vladimir Dogiel, Katia Ferriere, Nidhal Guessoum, Pierre Jean, J\"urgen Kn\"odlseder, Alexandre Marcowith, Pierrick Martin, Igor Moskalenko, Nikos Prantzos, Joe Silk, Andrew Strong, and Georg Weidenspointner.

%
\bibliographystyle{JHEP_MDL}
\providecommand{\href}[2]{#2}\begingroup\raggedright\endgroup


\end{document}